**Assessing the Impact of External and Internal Factors on Emergency Department Overcrowding**

Abdulaziz Ahmed[1,2*], Khalid Y. Aram[3], Mohammed Alzeen[1], Orhun Vural[4], James Booth[5], Brittany F. Lindsey[6], Bunyamin Ozaydin[1,2]

[1]Corresponding author, Department of Health Services Administration, School of Health Professions, University of Alabama at Birmingham, 1716 9th Ave S, Birmingham, AL 35233, USA (aahmed2@uab.edu)
[2]Department of Biomedical Informatics and Data Science, Heersink School of Medicine, University of Alabama at Birmingham, Birmingham, AL 35233, USA
[3]School of Business & Technology, Emporia State University, Emporia, KS 66801, United States
[4]Department of Electrical and Computer Engineering, University of Alabama at Birmingham, AL, 35233, USA
[5]Department of Emergency Medicine, University of Alabama at Birmingham, Birmingham, Alabama, 35233, USA
[6]Department of Patient Throughput, University of Alabama at Birmingham, Birmingham, AL, 35233, USA

**ABSTRACT**

**Study Objective**: To analyze the factors influencing Emergency Department (ED) overcrowding by examining the impacts of operational, environmental, and external variables, including weather conditions and football games.
**Methods**: This study integrates ED tracking and hospital census data from a southeastern U.S. academic medical center (2019–2023) with data from external sources, including weather, football events, and federal holidays. The dependent variable is the hourly waiting count in the ED. Seven regression models were developed to assess the effects of different predictors such as weather conditions, hospital census, federal holidays, and football games across different timestamps.
**Results:** Some weather conditions significantly increased ED crowding in the Baseline Model, while federal holidays and weekends consistently reduced waiting counts. Boarding count positively correlated with ED crowding when they are concurrent, but earlier boarding count (3-6 hours before) showed significant negative associations, reducing subsequent waiting counts. Hospital census exhibited a negative association in the Baseline Model but shifted to a positive effect in other models, reflecting its time-dependent influence on ED operations. Football games 12 hours before significantly increased waiting counts, while games 12 and 24 hours after had no significant effects.
**Conclusion**: This study highlights the importance of incorporating both operational and non-operational factors (e., weather) to understand ED patient flow. Identifying robust predictors such as weather, federal holidays, boarding count, and hospital census can inform dynamic resource allocation strategies to mitigate ED overcrowding effectively.

**Keywords**: Emergency Overcrowding, Weather and Emergency Visits, Football Game and Emergency Visits, Linear Regression, Quantitative Analysis.

INTRODUCTION

Emergency Departments (EDs) face a widespread crisis of overcrowding, with over 90% routinely operating beyond capacity during peak periods.[1] This systemic challenge, where emergency service demand exceeds available patient care resources, correlates with increased mortality. Studies demonstrate that when ED occupancy surpasses average levels, in-hospital mortality rates increase by approximately 3.1%, rising to 5.4% during peak overcrowding.[2] The problem manifests through extended wait times, hallway boarding, and overwhelmed staff, creating cascading effects throughout healthcare systems that compromise both patient safety and care quality.[3,4]

Various approaches to quantifying ED overcrowding have evolved over the past two decades, ranging from simple metrics to sophisticated scoring systems. The ED Occupancy Rate calculates the ratio of total patients to treatment beds,[5] while Length of Stay (LOS) for admitted patients serves as another key indicator of ED backup.[6] More sophisticated tools include the Emergency Department Work Index (EDWIN),[7] which incorporates five variables: number of ED patients in each triage category, patient triage category, number of treatment spaces in the ED, i.e., bays or beds, number of attending physicians currently on duty, and number of admitted patients in the ED, i.e., holds or inactive patients. The National Emergency Department Overcrowding Scale (NEDOCS)[8] incorporates seven variables, including total patients, bed capacity, ventilator use, and boarding times.[9] The Community Emergency Department Overcrowding Scale (CEDOCS)[10] builds upon NEDOCS by incorporating community hospital-specific variables, while the Severely Overcrowded, Overcrowded, and Not-overcrowded Estimation Tool (SONET)[11] introduces composite indices for total patient load and waiting room status. Despite these innovations, achieving adequate standardization across healthcare settings remains challenging.

ED overcrowding significantly impacts patient outcomes, correlating with increased mortality, longer hospital stays, and higher healthcare costs.[12] Asplin et al.'s framework[13] provides a comprehensive structure for understanding this crisis through three interconnected components: input, throughput, and output. Input factors encompass both operational elements that hospitals can directly manage and non-operational influences such as environmental conditions, temporal patterns, and community events.[14] Moreover, limited access to primary care and specialty services drives increased ED utilization,[15,16] while inadequate mental health resources direct crisis patients to EDs.[17] Additionally, insurance status significantly affects ED usage, particularly among populations with limited healthcare access alternatives.[18,19]

Throughput factors reflect internal ED efficiency, where workflow organization, staffing levels, and ancillary service delays create potential bottlenecks in patient processing.[20,21] The boarding of inpatients due to limited bed capacity significantly impacts throughput by consuming resources needed for new patients.[22] Recent developments in artificial intelligence and machine learning offer promising approaches for optimizing workflows and predicting patient flow.[23] Output factors are related to patient disposition, where organizational inefficiencies and delayed discharges create ripple effects throughout the hospital.[24] While existing research typically focuses on limited variables and narrow timeframes, the temporal relationships between factors remain understudied.[5,12]

Our study addresses these gaps by developing a comprehensive set of models analyzing factors affecting ED waiting room counts, a key overcrowding indicator.[25] Following Asplin's framework,[13,14] we examine three variable categories. Input factors include temporal elements (month, day, weekend status, hour) and external influences (sporting events, holidays, weather). Throughput factors encompass treatment count and real-time waiting room metrics, while output factors include boarding count and hospital census.

To capture temporal dynamics, we analyze these variables across multiple time lags, examining both immediate values and rolling averages. We incorporate lagged values of variables from 3 hours before to 24 hours after reference points. These temporal dimensions are crucial because while operational factors typically have immediate effects that hospitals can actively manage, non-operational factors often have delayed or sustained impacts requiring different management strategies. Further, our focus on external influences like weather conditions and sporting events is particularly significant, as prior research has shown that weather patterns can significantly affect ED utilization,[26] while mass gatherings like sporting events can create sudden surges in ED demand through both direct injuries and indirect effects on community behavior.[27,28]

**METHOD**

**Data sources**

Four main data sources were used to investigate factors that impact ED overcrowding: ED tracking data, inpatient data, weather information, and significant event dates. The ED tracking and inpatient data originated from an academic medical center located in the southeastern United States. Weather data was obtained from the OpenWeatherAPI[29], using data collected from a station near the hospital. Significant event dates include federal holidays and game times for a major local college football team in the nearby city. All four datasets cover the period from January 1, 2019, to July 1, 2023, and were merged on an hourly basis to create a comprehensive dataset for analysis.

The ED tracking dataset includes detailed records of patient visits, allowing tracking of patient movement within the ED. The dataset contains 161,477 unique patient IDs and 308,196 unique visit IDs. The dataset captures timestamps for ED arrival and discharge, as well as room codes that record when patients enter and exit various areas within the ED. Additionally, it includes an event column that records actions such as discharge, bed requests, triage, or physician exams throughout a patient's visit. The inpatient dataset includes records for all patients admitted to the hospital, containing 209,505 unique patient IDs. For each admission, the dataset includes admit and discharge timestamps. The weather dataset includes hourly environmental data collected using OpenWeatherAPI's history bulk. It contains numerical variables such as temperature, humidity, and wind speed, along with categorical weather statuses like clear or cloudy skies, rain, mist, thunderstorm, snow, drizzle, haze, fog, and smoke. Finally, the significant events dataset includes categorical variables for major events that could impact hospital activity, specifically local football games and federal holidays. Football game times were sourced from the team's official website,[30] with an average of 13 games occurring each year. Federal holiday dates were obtained from the U.S. government website,[31] covering 10 different holidays each year.

**Operationalization of Variables and Data Preprocessing**

We carried out several preprocessing steps to prepare the dataset for analysis. Each variable specified in Table 1 was aggregated on an hourly basis, with each row representing data for a specific one-hour interval. The variables waiting count, treatment count, and boarding count were derived from the ED tracking dataset. Waiting count, which is the dependent variable, represents the total number of patients who arrived at the ED at a specific hour but did not start treatment. Treatment count represents the total number of patients who are being treated in the ED at a specific hour. The boarding count represents the total number of patients whose treatments in the ED have been completed and who received admission orders from their emergency physician but remain in the ED while waiting for available beds in the inpatient units at any time within the hourly interval. The hospital census was derived from inpatient data, representing the total number of patients that are in inpatient units at any time within the hourly interval. Temperature, humidity, wind speed, and weather status were extracted from the weather dataset. Temperature and humidity were used to calculate the heat index using the heat index equation developed by Rothfusz (1990).[32] The Football Game and Federal Holiday variables were sourced from the "significant events" dataset. These preprocessed data were then merged based on the hourly format, resulting in a comprehensive dataset for analysis. All the variables were structured to be on an hourly basis. A descriptive analysis of the final dataset is provided in Table I.

We applied data-cleaning steps to enhance the quality of the data. The data from the COVID-19 period (from 03-01-2020 to 05-01-2021) was excluded, as hospital operations during this time were significantly different. Additionally, entries of patients who spent more than 9 hours in the waiting room, accounting for 1.89% of the entire dataset, were removed based on the recommendation of our clinical collaborators. The weather status feature was also simplified by grouping conditions into five categories: clear, clouds, rain, thunderstorm, and others (including fog, haze, snow, and smoke). Each weather status is represented as a binary variable: a value of 1 indicates that the specific weather condition is present in the row, while a value of 0 indicates it is not. Additional variables were created from some of the original variables listed in Table 1 by using their values 3-, 6-, 12-, and 24 hours before and after the current time of the dependent variable (See Table A1). This allows for capturing the temporal impact of those variables on overcrowding. For instance, the thunderstorm variables 3 and 6 hours before allow for an analysis of how these can be used to explain the effect of time-shifted thunderstorm statuses on the current value of the dependent variable. Lastly, rolling means, which calculate the average of a variable over a specific time window, were applied to hospital census, treatment count, and boarding to assess the impact of these variables' longer-term trends on overcrowding.

**Table 1: Summary of the Dataset**

| Feature | Ranges for Date/Time Variables, Average ± Standard Deviation (Range) for Numerical Variables, % for Categorical Variables, and Count for Event Variables |
|---|---|
| Month of Year | 1-12 Months |
| Day of Month | 1-31 Days |
| Day of Week | 1-7 Days |
| Hour of Day | 1-24 Hours |

| Waiting Count (Dependent Variable) | 18.23 ± 9.86 (0 – 59) |
|---|---|
| Treatment Count | 67.37 ± 23.52 (9 – 139) |
| Boarding Count | 45.72 ± 29.27 (3 – 121) |
| Hospital Census | 792.29 ± 71.66 (584 – 1017) |
| Heat Index | 86.2±24.9 |
| Wind Speed | 2.62 ± 2.05 m/s (0 – 15.4) |
| Weather Status | |
|     Clouds | 60.10% |
|     Clear | 22.78% |
|     Rain | 15.55% |
|     Thunderstorm | 1.25% |
|     Others | 0.33% |
| Football Game | 43 Games (across the entire date range) |
| Federal Holidays | 37 Days (across the entire date range) |

**Outcomes**

The outcome variable in this study is waiting count, representing the ED crowding. The waiting count is the total number of patients in the waiting room at a specific one-hour interval. Patients were included in an hourly count if their waiting room arrival was before the end of the interval and their departure was after its beginning. The total count for each hourly interval was determined by counting all patients who met these criteria. For instance, to calculate the total number of patients at the 1:00 pm to 2:00 pm interval, any patient who arrives at the ED before 2:00 pm and departs anytime at or after 1:00 pm was counted for the interval. In another instance, a patient who arrives at 12:50 PM and departs at 2:10 PM was counted in the intervals 12–1 PM, 1–2 PM, and 2–3 PM.

**Analysis**

Statistical analyses were conducted in two steps. First, a preliminary analysis was performed to examine how individual independent variables influence the dependent variable (i.e., the waiting count). The primary objective of this analysis is to understand the distribution and impact of each variable on the dependent variable, providing a foundational understanding before moving to multivariate approaches. In the second step, multivariate linear regression was conducted to analyze the effects of the independent variables on the waiting count. This analysis aims to quantify the impact of various independent variables, including football game events, weather, and downstream ED metrics such as patient treatment and boarding count on the number of waiting patients.

**Preliminary analysis**

We started this analysis by inspecting the distributions of the numerical independent variables using histograms. Then, we explored the relationships between these variables, such as hospital census and treatment count, with the dependent variable using t-tests. This analysis enabled us to identify trends, correlations, and potential outliers. For categorical variables, such as weather status, federal holidays, and football game events, we used t-tests to assess the statistical significance of the difference in the waiting count between different categories.

Additionally, we used confidence interval plots to illustrate the difference in the dependent variable across these categories, highlighting the direction and magnitude of the difference. The preliminary analysis provides initial insights into the variables that may contribute to the waiting count, setting the stage for subsequent multivariate analysis. Square root transformation was applied to the dependent variable to achieve normally distributed residuals.

**Multivariate Regression Analysis**

To further investigate the relationship between the dependent variable (hourly waiting count) and independent variables (downstream metrics and external factors such as weather), seven multivariate linear regression models were built. The models were designed to investigate the effects of independent variables at different timestamps on the dependent variable. The model design included a baseline set of variables, then extended to models that incorporate different timeframes before and after the current value of the independent variables to assess the impact of temporal changes on waiting count values, ensuring the robustness of the proposed regression analysis. In the **Baseline Model**, concurrent independent variables such as football game actual time, hospital census, treatment count, and weather statuses like clear, clouds, rain, and thunderstorm were included. This served as a foundation for understanding how these independent variables relate to the dependent variable.

To further enhance the robustness of the analysis and capture temporal variations, we introduced additional models that incorporate variables reflecting different time frames: **Model 2: 3 Hours Before** builds on the baseline by retaining key variables such as month, day of month, hour, day of week, heat index, and wind speed, while including additional features such as weather conditions (clear, clouds, rain, thunderstorm), hospital census data, and treatment and boarding counts to three hours before the time of the target event. In other words, **Model 2: 3 Hours Before** explains the effects of those variables $t - 3$, on the dependent variable at $t$ Additionally, football game time was shifted to 12 hours before the target event.

**Model 3: 6 Hours Before** retains the same dependent variables as Model 2 but uses 6-hour early values ($t - 6$ values), except for the football game variable, which uses 24 hours before the target event. This design ensured that any significant trends manifesting over a longer time period were captured, hence enhancing the model's robustness. **Model 4: 12 Hours After** and **Model 5: 24 Hours After** include similar sets of variables to the base model, except that model 4 considers a 12-hour timeframe for the football game, while model 5 considers 24 hours after. The inclusion of variables reflecting post-event dynamics provides valuable insights into how these variables influence patient flow beyond immediate effects.

**Models 6 and 7** introduce rolling averages for key variables such as treatment count and hospital census data. Models 6 and 7 include treatment and boarding counts calculated over 6 hours window. Model 6 includes hospital census calculated over 12 hours, while model 7 includes hospital census calculated over 24 hours. These rolling averages capture cumulative effects over time, allowing the model to consider not only instantaneous impacts but also the influence of trends over extended periods.

Incorporating variables across different timeframes results in a comprehensive and robust regression analysis. This design ensures that the model can generalize effectively to different scenarios, considering the dynamics of patient flow in the waiting room under varying temporal and environmental conditions.

**RESULTS**

**Preliminary Analysis Results**

We analyzed the effects of the categorical variables in the data on the dependent variable (i.e., waiting count). Figure 1 shows confidence interval plots for the mean difference in waiting count between the categories of the variables. The categorical variables in the data are all binary with the following two categories: 1) Yes, which indicates the presence of the event, such as a Federal Holiday, and 2) No, which indicates the absence of the event. Figure 1 also provides the p-values of t-tests that were used to assess the statistical significance of the mean differences displayed in the figure. All the t-test tests were performed with $\alpha = 0.05$ and assuming equal population variances. Football Game times seem to influence the waiting count, particularly the games that happened 24 hours before (figure 1-B) or will happen 12 hours later (figure 1-E). The t-test for the game happening 12 hours after indicates a significant decrease in current ED overcrowding (mean difference $\approx$ -6, p-value = 0.0006). On the contrary, the test shows a significant increase in the waiting count (mean difference $\approx$ 6.3, p-value = 0.0008) for games that happened 24 hours earlier. Further, the t-tests show that football games that happened at the actual time of the recorded waiting count, 12 hours before and 24 hours after, do not have significant effects on waiting counts.

Federal Holidays resulted in a statistically significant decrease in ED waiting count (figure 1-F). The results also show a significant effect of weather on ED overcrowding. Clear weather is associated with a statistically significant decrease (mean difference $\approx$ -1.2) in waiting counts (figure 1-G). In contrast, cloudy weather correlates with a statistically significant increase (mean difference $\approx$ 0.5) in waiting count (figure 1-H). Rain and thunderstorms show an association with a statistically significant increase in waiting count (mean differences $\approx$ 0.8 and 3, respectively) as well (figures 1-J and 1-K). The "Other" weather variable shows an association with a significant increase in waiting counts. The other weather category includes drizzle, fog, haze, snow, and smoke, but in small percentages.

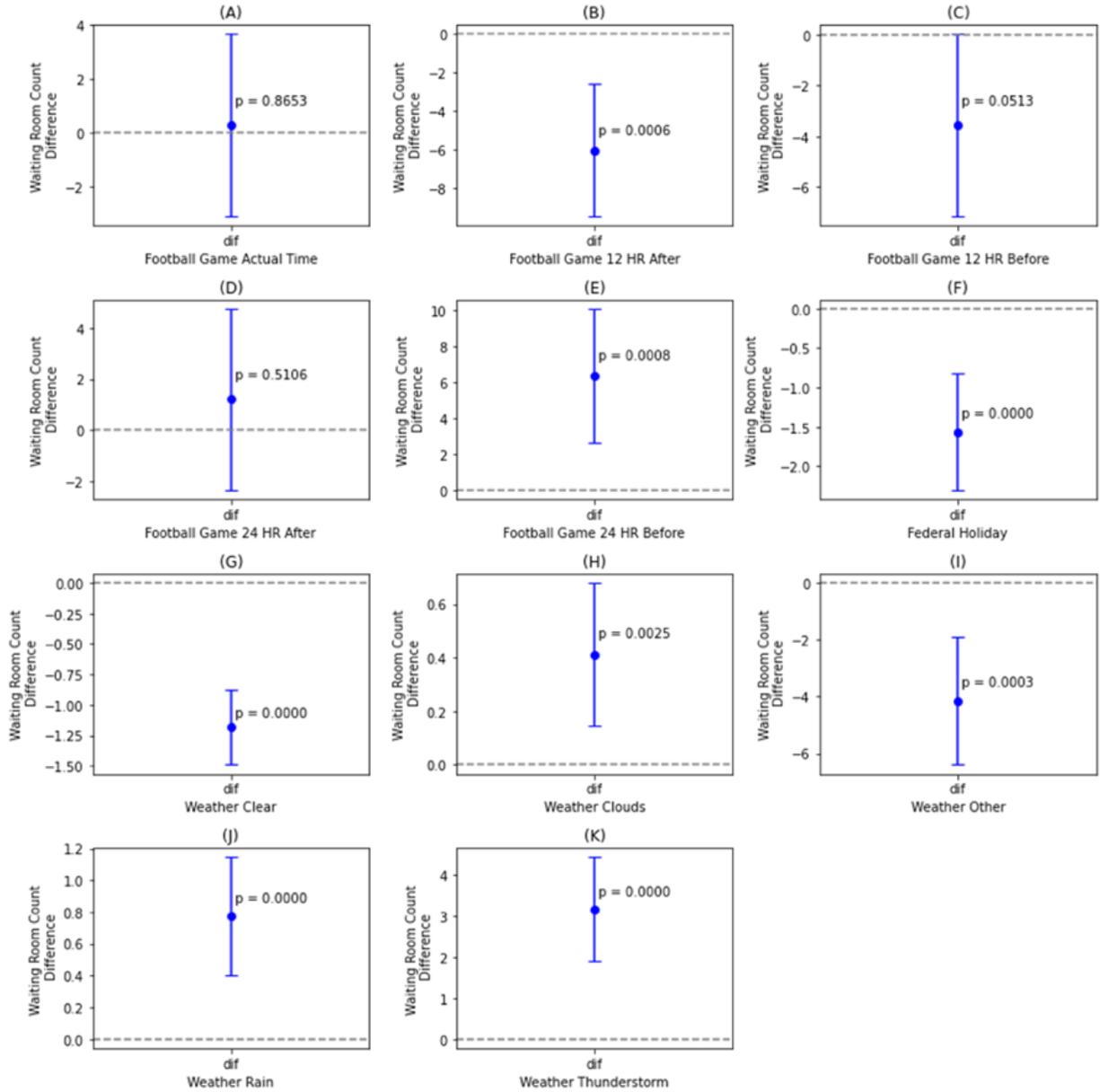

*Figure* 1*: Confidence interval plots of ED waiting count with respect to football game, federal holiday, and weather variables.*

**Multivariate Regression Results**

The multivariate regression results presented in Table 2 provide insights into how various factors influence the total waiting count and how these effects evolve with the timing of certain independent variables, such as football games and weather conditions. This underscores the robustness of the model design. To fulfill the normality assumption of linear regression (See Figure A1 in the appendix), we applied square root transformation to the dependent variable. The regression coefficients were used to interpret the impact of the independent variables on the square root of waiting count, meaning that for each unit of change in a certain independent variable, the square root of the waiting count changes by the corresponding coefficient.

In the Baseline Model, weather variables exhibit significant positive associations with varying degrees of impact on the square root of waiting count. For example, weather clear (1.058***), weather clouds (1.064***), weather rain (1.004***), and weather thunderstorm (1.182***) indicate that all weather conditions are associated with an increase in waiting counts and consequently worsen ED overcrowding, however, with varying degrees. Among weather categories, clear, cloudy, and rainy weather have close effects, but thunderstorms had the highest effect on ED overcrowding. While wind speed (0.042***) causes an increase in waiting count, heat index (-0.003***) is associated with a smaller waiting count. Federal Holiday (-0.636***) and weekend (-0.800***) significantly reduce waiting counts, indicating that holidays and weekends are associated with fewer patients in the waiting room and consequently a less crowded ED. Boarding count (0.002***), which is a downstream independent variable, is positively associated with waiting counts, while hospital census (-0.003***) shows a significant negative relationship. Temporal variables such as hour (0.072***), month (0.004*), and day of the month (-0.005***) capture seasonal and hourly variations in patient flow.

Model 2 incorporates variables that were measured three hours before the target time. Weather Clear 3 Hours Before (-0.245), Weather Clouds 3 Hours Before (-0.143), Weather Rain 3 Hours Before (-0.206), and weather thunderstorm 3 hours before exhibited negative coefficients, yet none was significant. This result implies that the weather 3 hours before does not have a significant effect on the current waiting counts, consequently the ED crowding. Wind Speed 3 Hours Before (0.049***) remained positively associated with increased waiting counts. Federal Holiday (-0.448***) and Weekend (-0.624***) remained significant, with their coefficients slightly reduced compared to the Baseline Model. Hospital Census 3 Hours Before (0.001***) retains its significance but became positively associated with waiting count compared to the Base Model (-0.003***). Boarding Count 3 Hours Before (-0.002***) remained significant compared to the Baseline Model (0.002***). Football Game 12 Hours Before had a significant impact (0.305*) on waiting count compared to the football game at the current time in the Base Model (0.104), which was insignificant.

In Model 3, weather variables measured 6 hours before the current time show stronger negative impacts than their effects in Model 2, but they are still insignificant. Federal Holiday (-0.365***) and weekend (-0.554***) continue to have significant negative effects, consistent with earlier models, despite being less profound. Wind Speed 6 Hours Before (0.034***) remained significant and positive. Boarding Count 6 Hours Before (-0.004***) is a similar effect to 3 hours before in Model 2 (-0.002). Hospital Census 6 Hours Before (0.003***) remained significant, but its impact increased compared to the 3 hours before in model 2 (0.001***). The Football Game 24 Hours Before (0.207) is also associated with an increase in ED waiting count but is not statistically significant, compared to the 12 hours in Model 2 (0.305*), where it is marginally significant.

Models 4 and 5 evaluate the effects of football games 12 and 24 hours after the dependent variable time compared to the Baseline Model, respectively. In both Model 4 and 5, federal holiday (-0.643*** in both models) and weekend (-0.799*** and -0.800*** in models 4 and 5 respectively) maintained their significant negative impacts on waiting counts, reflecting their consistent influence across models. Weather conditions also maintained significant, positive associations and similar impact compared to the Baseline Model. Meanwhile, the effects of football games 12 hours after (-0.122) became statistically insignificant in contrast with the 12 hours before in Model 2, which was significant. Football games 24 hours after (0.146) effect was consistent with that of football games 24 before in Model 3 (0.207), in which both were insignificant.

Models 6 and 7, compared to the Baseline Model, introduce rolling averages for hospital census and boarding count, capturing their cumulative effects over time. In Model 6, the coefficients of rolling mean boarding count window size 6 (-0.003***) and rolling mean hospital census window size 12 (0.002***) are statistically significant. While the rolling mean of the boarding count coefficient became negative compared to its coefficient in the Baseline Model (0.002***), the rolling mean for hospital census became positively associated with waiting count compared to the negative impact of its non-rolling average version in the Baseline Model (-0.003***). This means that the increase in hospital census (i.e., rolling mean) in previous hours increased the waiting count in ED. In Model 7, both the rolling mean boarding count window size 6 (-0.001*) and hospital census window size 24 (-0.000) are negatively associated with waiting count, but the effect is minimal. The rolling mean of boarding count became negatively associated with the waiting count compared to its significant positive association in the Baseline Model (0.002***). Comparing Models 6 and 7, where rolling mean for hospital census considered past 12 hours (Model 6), as opposed to 24 (Model 7), the direction of impact remained the same, but less significant for the rolling mean of the boarding count, whereas the direction of impact switched for hospital census from positive to negative with decreased non-significant impact.

The R-squared ($R^2$) and Adjusted $R^2$ values indicate how well each model explains the variance in the square root of the waiting count. The Baseline Model had an $R^2$ of 0.285, explaining 28.5% of the variation, with a similarly strong Adjusted $R^2$ of 0.285. Model 2 (3 hours before) showed a higher $R^2$ of 0.290, indicating a stronger predictive power, while Model 3 (6 hours before) showed a higher $R^2$ of 0.306. In contrast, both Model 4 (12 hours after) and Model 5 (24 Hours After) showed $R^2$ of 0.285, lower than Model 3 and similar to the Baseline Model. Models 6 and 7, which include rolling averages, explain 38.9% and 38.2% of the variance, respectively.

Table 12: 2Regression results for all models.

| | Baseline Model | Model 2: 3 Hours Before | Model 3: 6 Hours Before | Model 4: 12 Hours After | Model 5: 24 Hours After | Model 6: Rolling Averages | Model 7: Rolling Averages 2 |
|---|---|---|---|---|---|---|---|
| Football Game Actual Time | 0.104 | | | | | 0.140 | 0.134 |
| Hospital Census | -0.003(***) | | | -0.003(***) | -0.003(***) | | |
| Boarding Count | 0.002(***) | | | 0.002(***) | 0.002(***) | | |
| Federal Holiday | -0.636(***) | -0.448(***) | -0.365(***) | -0.635(***) | -0.635(***) | -0.409(***) | -0.509(***) |
| Weather Clear | 1.058(***) | | | 1.059(***) | 1.058(***) | 0.512(***) | 0.757(***) |
| Weather Clouds | 1.064(***) | | | 1.064(***) | 1.064(***) | 0.486(***) | 0.740(***) |
| Weather Rain | 1.004(***) | | | 1.004(***) | 1.003(***) | 0.449(***) | 0.691(***) |
| Weather Thunderstorm | 1.182(***) | | | 1.181(***) | 1.181(***) | 0.605(***) | 0.877(***) |
| Heat Index | -0.003(***) | -0.002(***) | -0.001(***) | -0.003(***) | -0.003(***) | -0.002(***) | -0.002(***) |
| Wind Speed | 0.042(***) | | | 0.042(***) | 0.042(***) | 0.038(***) | 0.040(***) |
| Weather Others | 0.574(***) | | | 0.575(***) | 0.575(***) | -0.012 | 0.227(**) |
| Month of year | 0.004(*) | 0.001 | -0.002 | 0.004(*) | 0.004(*) | -0.002 | 0.001 |
| Day Of Month | -0.005(***) | -0.004(***) | -0.004(***) | -0.005(***) | -0.005(***) | -0.004(***) | -0.005(***) |
| Hour of day | 0.072(***) | 0.073(***) | 0.067(***) | 0.072(***) | 0.072(***) | 0.074(***) | 0.076(***) |
| Weekend | -0.800(***) | -0.624(***) | -0.554(***) | -0.799(***) | -0.800(***) | -0.598(***) | -0.677(***) |
| Weather Clear 3 Hours Before | | -0.245 | | | | | |

| | | | | | | |
|---|---|---|---|---|---|---|
| Weather Clouds 3 Hours Before | | -0.143 | | | | |
| Weather Rain 3 Hours Before | | -0.206 | | | | |
| Weather Thunderstorm 3 Hours Before | | -0.188 | | | | |
| Weather Other Hours Before | | -0.411 | | | | |
| Wind Speed 3 Hours Before | | 0.049(***) | | | | |
| Boarding Count 3 Hours Before | | -0.002(***) | | | | |
| Hospital Census 3 Hours Before | | 0.001(***) | | | | |
| Football Game 12 Hours Before | | 0.305(*) | | | | |
| Football Game 12 Hours After | | | | -0.122 | | |
| Weather Clear 6 Hours Before | | | -0.736 | | | |
| Weather Clouds 6 Hours Before | | | -0.566 | | | |
| Weather Rain 6 Hours Before | | | -0.646 | | | |
| Weather Thunderstorm 6 Hours Before | | | -0.549 | | | |
| Weather Other Hours Before | | | -0.549 | | | |
| Wind Speed 6 Hours Before | | | 0.034(***) | | | |
| Boarding Count 6 Hours Before | | | -0.004(***) | | | |
| Hospital Census 6 Hours Before | | | 0.003(***) | | | |
| Football Game 24 Hours Before | | | 0.207 | | | |
| Football Game 24 Hours After | | | | 0.146 | | |
| Rolling Mean Boarding Count Window Size 6 | | | | | -0.003(***) | -0.001(*) |
| Rolling Mean Hospital Census Window Size 12 | | | | | 0.002(***) | |
| Rolling Mean Hospital Census Window Size 24 | | | | | | -0.000 |
| *p<0.05, **p<0.01, ***p<0.001 | | | | | | |

## DISCUSSION

The proposed regression models provided insight into ED waiting room overcrowding, emphasizing the importance of temporal and cumulative effects of certain metrics in understanding patient flow. The models in this study offer a comprehensive view of interesting variables that influence ED congestion. The models also focused on the temporal effects of these variables by considering their values over varying time windows. The regression results demonstrated that weather conditions play a significant role in influencing ED waiting counts. In the Baseline Model, weather categories—clear, cloudy, rainy, and thunderstorms—exhibited significant positive associations with waiting counts, with thunderstorms having the strongest impact. This indicates that weather conditions, particularly thunderstorms and wind speed, may lead to increased ED visits due to factors such as accidents or exacerbations of chronic illnesses. This is consistent with t-test results, which also showed a higher waiting count during thunderstorms, highlighting the immediate strain that severe weather conditions place on ED resources. While wind speed causes an increase in waiting count, the heat index shows a small but statistically significant negative relationship, likely reflecting reduced outdoor activity during extreme heat. When weather conditions three and six hours before the target time are used (Models 2 and 3), their effects became statistically insignificant, indicating that the influence of prior weather dissipates over time. The consistent significance of wind speed across all models highlights its robust and persistent role in influencing ED overcrowding. Overall, weather conditions directly impact ED overcrowding, with thunderstorms and wind speed contributing most significantly. These weather impacts are consistent with recent studies showing a 1.10 IRR increase in ED visits during rain and 1.07 IRR during snow events, with lower utilization on weekends and nights.[33]

The impact of football games on ED waiting counts appears to be influenced by the timing of the event relative to patient flow. While the t-test results suggest significant changes in waiting counts for games occurring 24 hours prior and 12 hours after, these associations were not supported by the regression models. In fact, the regression analysis showed that only games occurring 12 hours before the target time had a significant positive association with waiting counts, while games at the actual time, 24 hours before, and 12 or 24 hours after were not significant across the models. This discrepancy highlights the limitations of bivariate analyses like t-tests, which do not account for confounding variables such as hospital census, boarding counts, time of day, and weather conditions. In contrast, the regression models control for these factors, offering a more comprehensive understanding of the dynamics at play. The significant association observed in the regression for games occurring 12 hours before the target time is particularly noteworthy, as it likely reflects a more valid relationship between football events and ED overcrowding. This suggests that the increased waiting counts seen around these events are not solely due to the games themselves but are also shaped by underlying operational and temporal factors. The lack of significant effects for games 24 hours before or after in the regression, despite their significance in the t-tests, further underscores the importance of multivariate analyses in identifying true associations amidst complex healthcare environments. Hence, regression-based findings provide a stronger basis for understanding how external events like football games influence ED crowding, emphasizing the need for comprehensive models when evaluating such multifaceted relationships. These findings are further supported by Antowiak et al. (2021), where their analysis showed modest 2.6% reductions in ED volume during regional football games and 4.3% during Monday night games.[34]

Hospital Census demonstrates a complex, time-dependent relationship with ED waiting counts. In the Baseline Model, hospital census exhibited a significant negative association with ED waiting counts, which might initially appear counterintuitive. This negative relationship likely reflects a lagged effect, where high inpatient occupancy eventually limits ED throughput and indirectly affects waiting counts. In Models 2 and 3, when the hospital census is measured three and six hours before the target time, its effect shifted to a significant positive association with waiting counts. This indicates that earlier increases in inpatient census exacerbate ED overcrowding over time. Rolling averages in Models 6 and 7 further highlight the cumulative impact of hospital census. In Model 6, the rolling mean of hospital census over 12 hours was positively associated with waiting counts, indicating that prolonged high inpatient occupancy increases ED overcrowding. However, in Model 7, the rolling mean over 24 hours remained significant but with minimal effect, reflecting a diminishing cumulative effect of hospital census over extended periods. These findings emphasize the temporal complexity of hospital census and its lagged influence on ED dynamics.

Boarding count was found to be a significant variable influencing ED overcrowding, with its effects varying across models. In the Baseline Model, boarding count is positively associated with waiting count, indicating that higher boarding counts are correlated with increased ED overcrowding. This reflects the direct impact of accommodating patients awaiting inpatient beds, which adds to the strain on ED resources and slows patient throughput. However, in Model 2, the boarding count measured three hours earlier shows a significant negative association with waiting count. This counterintuitive finding may reflect temporal patterns in patient flow rather than a direct causal effect. One possible explanation is that boarding counts tend to rise following periods of high patient volumes, which often occur earlier in the day. By the time boarding counts increase, waiting counts may have naturally decreased due to lower patient arrivals during off-peak hours (e.g., late-night shifts). In this context, waiting counts might influence boarding counts more than the reverse. In Model 3, the negative association becomes more pronounced with boarding counts measured six hours earlier, potentially reflecting compounded temporal effects.

The rolling averages in Models 6 and 7 reveal patterns in the association between boarding counts and ED waiting counts over time. In Model 6, the rolling mean of boarding count over 6 hours shows a significant negative association with ED waiting counts. In Model 7, the rolling mean over 24 hours also exhibits a negative association, though the effect is smaller. These results suggest that the relationship between boarding activity and waiting counts may vary over different timeframes. However, these associations should not be interpreted as causal. It is possible that both boarding and waiting counts are influenced by underlying temporal factors, such as time-of-day effects or patient flow patterns. While higher real-time boarding counts are associated with increased waiting room congestion, the negative associations observed with earlier boarding activity may reflect temporal dynamics rather than direct effects. Further analysis is needed to clarify these relationships.

Federal holidays and weekends consistently exhibit significant negative associations with waiting counts across all models, reflecting reduced patient volumes during these periods. These findings are consistent over the models and reinforce the idea that ED overcrowding decreases on holidays and weekends due to altered hospital operations or reduced healthcare demand. Time-based variables, hour of day, month year, and day of the month, capture predictable seasonal and hourly variations in patient flow. The hour variable consistently shows a significant positive association with waiting counts. This indicates these variables contribute to ED crowding. The Month variable exhibits a smaller but significant positive effect, indicating variations in patient volumes throughout a given month, while the day of the month is negatively associated with waiting counts, though its impact is minimal.

While several variables show statistically significant associations with ED waiting counts across models, it is important to note that the coefficients for some of these variables are relatively small. For example, variables such as boarding count and hospital census, despite their statistical significance, exhibit coefficients that indicate minimal practical impact on waiting counts. This suggests that while these factors are consistently associated with ED overcrowding, their individual effects may be limited in magnitude. These small coefficients may reflect the complex, multifactorial nature of ED dynamics, where no single factor overwhelmingly drives overcrowding. Therefore, the clinical or operational relevance of these variables should be interpreted in the context of the broader system, where cumulative effects of multiple small influences can still be meaningful. Further research could explore interactions between these variables to better understand their combined impact on ED crowding.

**LIMITATION**

This study, while offering valuable insights into ED overcrowding, had multiple limitations. First, the analysis was based on data from a single southeastern U.S. academic medical center. This might limit the generalizability of findings to other regions or healthcare systems due to varying patient demographics, workflows, policies and regulations, or environmental conditions. The regression models rely on linear relationships and temporal shifts, which may not capture complex, non-linear interactions among variables. Advanced machine learning approaches could provide deeper insights into these dynamics. Finally, preprocessing steps such as excluding patients waiting more than nine hours may inadvertently omit extreme cases that could provide valuable insights into severe overcrowding scenarios.

## CONCLUSIONS

The regression results highlight the multifaceted and time-dependent nature of ED overcrowding. Weather conditions, particularly thunderstorms and wind speed, are significantly associated with overcrowding, while football games influence waiting counts primarily 12 hours before the event. Hospital census and boarding count exhibit complex, lagged effects, with cumulative metrics providing additional insights into their sustained impacts. Federal holidays and weekends consistently reduce overcrowding, while temporal variables capture predictable variations in patient flow. It is also noteworthy that some variables, despite being statistically significant, have relatively small coefficients, suggesting limited practical impact when considered individually. These findings highlight the importance of incorporating temporal and cumulative metrics into ED management strategies and considering the combined effects of multiple factors to dynamically allocate resources and mitigate overcrowding effectively.

## FUNDING INFORMATION
This project was supported by the Agency for Healthcare Research and Quality (AHRQ) under grant number 1R21HS029410-01A1.

## CONFLICT OF INTEREST STATEMENT
The authors declare no conflicts of interest.

## DATA AVAILABILITY STATEMENT
Data supporting the findings of this study are available upon reasonable request and subject to approval by the Institutional Review Board (IRB).

4

# APPENDIX

*Table A 1: Model design*

| | Baseline Model | Model 2: 3 Hours Before | Model 3: 6 Hours Before | Model 4: 12 Hours After | Model 5: 24 Hours After | Model 6: Rolling Averages | Model 7: Rolling Averages 2 |
|---|---|---|---|---|---|---|---|
| Football Game Actual Time | x | | | | | x | x |
| Hospital Census | x | | | x | x | | |
| Boarding Count | x | | | x | x | | |
| Federal Holiday | x | x | x | x | x | x | x |
| Weather Clear | x | | | x | x | x | x |
| Weather Clouds | x | | | x | x | x | x |
| Weather Rain | x | | | x | x | x | x |
| Weather Thunderstorm | x | | | x | x | x | x |
| Weather Others | x | | | x | x | x | x |
| Month | x | x | x | x | x | x | x |
| Day Of Month | x | x | x | x | x | x | x |
| Hour | x | x | x | x | x | x | x |
| Day Of Week | x | x | x | x | x | x | x |
| Heat Index | x | x | x | x | x | x | x |
| Wind Speed | x | | | x | x | x | x |
| Weather Clear 3 Hours Before | | x | | | | | |
| Weather Clouds 3 Hours Before | | x | | | | | |
| Weather Rain 3 Hours Before | | x | | | | | |
| Weather Thunderstorm 3 Hours Before | | x | | | | | |
| Weather Others 3 Hours Before | | x | | | | | |
| Wind Speed 3 Hours Before | | x | | | | | |
| Treatment Count 3 Hours Before | | x | | | | | |
| Boarding Count 3 Hours Before | | x | | | | | |
| Hospital Census 3 Hours Before | | x | | | | | |
| Football Game 12 Hours Before | | x | | | | | |
| Football Game 12 Hours After | | | | x | | | |
| Weather Clear 6 Hours Before | | | x | | | | |
| Weather Clouds 6 Hours Before | | | x | | | | |
| Weather Rain 6 Hours Before | | | x | | | | |

| | | | | | | |
|---|---|---|---|---|---|---|
| Weather Thunderstorm 6 Hours Before | | x | | | | |
| Weather Others 6 Hours Before | | x | | | | |
| Wind Speed 6 Hours Before | | x | | | | |
| Treatment Count 6 Hours Before | | x | | | | |
| Boarding Count 6 Hours Before | | x | | | | |
| Hospital Census 6 Hours Before | | x | | | | |
| Football Game 24 Hours Before | | x | | | | |
| Football Game 24 Hours After | | | | x | | |
| Rolling Mean Treatment Count Window Size 6 | | | | | x | x |
| Rolling Mean Boarding Count Window Size 6 | | | | | x | x |
| Rolling Mean Hospital Census Window Size 12 | | | | | x | |
| Rolling Mean Hospital Census Window Size 24 | | | | | | x |

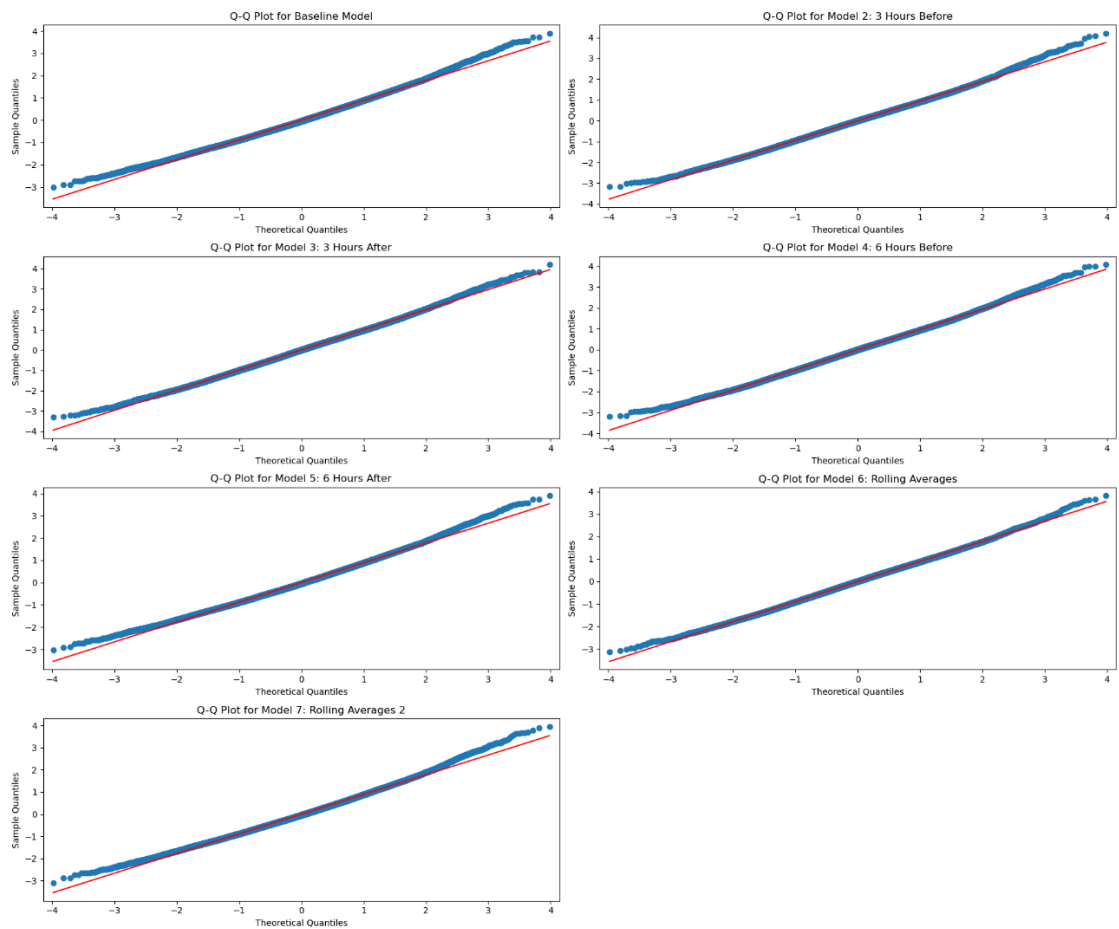

*Figure A 1: Normality test for the residuals of all models.*